\documentclass[a4paper,11pt]{article}
\pdfoutput=1 

\usepackage{jcappub} 

\usepackage[T1]{fontenc} 

\newcommand{\be}{\begin{equation}}
	\newcommand{\ee}{\end{equation}}
\newcommand{\ba}{\begin{eqnarray}}
	\newcommand{\ea}{\end{eqnarray}}

\def\l{\left}
\def\r{\right}

\newcommand{\gsim}{\mathrel{\hbox{\rlap{\lower.55ex \hbox {$\sim$}}
			\kern-.3em \raise.4ex \hbox{$>$}}}}
\newcommand{\lsim}{\mathrel{\hbox{\rlap{\lower.55ex \hbox {$\sim$}}
			\kern-.3em \raise.4ex \hbox{$<$}}}}

\title{\boldmath Primordial black holes from sound speed resonance in the inflaton-curvaton mixed scenario}

\author[a,b,c]{Chao Chen,}
\author[a,b,c,*]{and Yi-Fu Cai}

\affiliation[a]{Department of Astronomy, School of Physical Sciences, University of Science and Technology of China, Hefei, Anhui 230026, China}
\affiliation[b]{CAS Key Laboratory for Researches in Galaxies and Cosmology, University of Science and Technology of China, Hefei, Anhui 230026, China}
\affiliation[c]{School of Astronomy and Space Science, University of Science and Technology of China, Hefei, Anhui 230026, China}

\emailAdd{cchao012@mail.ustc.edu.cn}
\emailAdd{yifucai@ustc.edu.cn}

\abstract{We study sound speed resonance (SSR) mechanism for primordial black hole (PBH) formation in an early universe scenario with inflaton and curvaton being mixed. In this scenario, the total primordial density perturbations could be contributed by the fluctuations from both the inflaton and curvaton fields, in which the inflaton fluctuations lead to the standard adiabatic perturbations, while the sound speed of the curvaton fluctuations are assumed to be oscillating during inflation. Due to the narrow resonance effect of SSR mechanism, we acquire the enhanced primordial density perturbations on small scales and they remain nearly scale-invariant on large scales, which is essential for PBH formation. Finally, we find that the PBHs with specific mass spectrum can be produced with a sufficient abundance for dark matter in the mixed scenario.}

\keywords{primordial black holes, sound speed resonance, inflatonary cosmology, curvaton}

\arxivnumber{1908.03942}

\begin{document}
\maketitle
\flushbottom

\section{Introduction}
\label{sec:intro}

Primordial black holes (PBHs) may be formed from density fluctuations in the very early Universe \cite{Zeldovich:1966, Hawking:1971ei, Carr:1974nx}, which can be tested through their effects on a variety of cosmological and astronomical processes. In this regard, the PBHs can serve as an inspiring tool to probe physics in the very early Universe \cite{Khlopov:2008qy, Sasaki:2018dmp}. In particular, PBH could be a potential candidate for (a fraction of) dark matter (DM), which has drawn a lot of attention \cite{Carr:2016drx, Carr:2018poi}. PBHs with masses $\lesssim 10^{15}$ g would have evaporated at the present time due to the emission of Hawking radiation \cite{Hawking:1974rv}, and the emitted particles may impact the gamma-ray background \cite{MacGibbon:1991vc} and the abundance of light elements produced by the big bang nucleosynthesis \cite{Carr:2009jm}. For the PBHs with masses greater than $10^{15}$g, they could survive up until the present epoch and are expected to be constrained by their gravitational effects, such as gravitational lensing \cite{Niikura:2019kqi}, dynamical effects on baryonic matter \cite{Carr:1997cn}, or the 
fast radio burst created by mergers of charged PBHs \cite{Deng:2018wmy}. Recently, particular attention has been paid to the gravitational waves (GWs) generated by PBH-PBH mergers \cite{Sasaki:2016jop, Mandic:2016lcn, Wang:2016ana}, as well as the induced GWs from the enhanced primordial density perturbations associated with PBH formation \cite{Baumann:2007zm, Ananda:2006af, Kohri:2018awv, Bartolo:2018rku, Cai:2019jah, Cai:2018dig}. The GW survey shall be a promising window to reveal physical processes of PBH formations.

Some high-density regions of the very early Universe are expected for PBH formation. One possibility is that there were large primordial inhomogeneities, and the resulting over-dense regions might collapse to form PBHs \cite{Carr:1975qj}. This motivates many theoretical mechanisms of generating PBHs, which often require a power spectrum of primordial density perturbations to be suitably large on certain scales that are associated with a particularly tuned background dynamics of quantum fields in the very early Universe (e.g. see \cite{GarciaBellido:1996qt, Garcia-Bellido:2017mdw, Domcke:2017fix, Kannike:2017bxn, Carr:2017edp, Ballesteros:2017fsr, Hertzberg:2017dkh, Franciolini:2018vbk, Kohri:2018qtx, Ozsoy:2018flq, Biagetti:2018pjj, Ballesteros:2018wlw, Georg:2019jld, Pi:2019ihn, Dalianis:2019asr, Wen:2019kxv, Fu:2019ttf, Motohashi:2017kbs, Liu:2019lul, Carr:2018poi, Germani:2018jgr, Kamenshchik:2018sig} for studies within inflation, see \cite{Chen:2016kjx, Quintin:2016qro} for discussions within bounce cosmologies, and see \cite{Sasaki:2018dmp} for recent comprehensive reviews).

Recently, a novel mechanism for PBH formation by virtue of sound speed resonance (SSR) was proposed in \cite{Cai:2018tuh}. According to this mechanism, it was found that an oscillating sound speed square could yield non-perturbative parametric amplification on certain perturbation modes during inflation. As a result, the power spectrum of primordial density perturbations can have a narrow major peak on small scales, while remains nearly scale-invariant on large scales as predicted by inflationary cosmology. Note that several minor peaks of the power spectrum of primordial density perturbations on smaller scales are also predicted by this mechanism. It was found in \cite{Cai:2018tuh} that the formation of PBHs caused by the resulting peaks in SSR mechanism could be very efficient, which could be testable in the future observational experiments. 

Primordial density perturbations seeded the large-scale structure (LSS) of the Universe, are usually thought to arise from the quantum fluctuations during the inflationary era, from which a nearly scale-invariant power spectrum with a standard dispersion relation is obtained \cite{Mukhanov:1990me}. This was confirmed by various cosmological measurements such as the cosmic microwave background (CMB) radiation \cite{Ade:2015lrj, Akrami:2018odb} and LSS surveys at extremely high precision. It is interesting to note that, any light scalar field during inflation could lead to a nearly scale-invariant power spectrum. This leads to an alternative mechanism for generating nearly scale-invariant primordial power spectrum from an additional light scalar field rather than the inflaton, which is called the curvaton scenario \cite{Linde:1996gt, Lyth:2001nq, Enqvist:2001zp, Lyth:2002my, Moroi:2001ct} (see also \cite{Cai:2011zx, Cai:2011ci, Cai:2013kja, Cai:2014bea, Alexander:2014uaa, Addazi:2016rnz} for discussions within bounce cosmologies), and it can be further generalized by including both inflaton and curvaton, which is dubbed as the  inflaton-curvaton mixed scenario \cite{Langlois:2004nn, Ferrer:2004nv, Langlois:2008vk, Cai:2009hw, Fonseca:2012cj, Byrnes:2014xua, Jiang:2018uce}. In the latter case, primordial density perturbations are generated by both the inflaton and curvaton fluctuations, in which the inflaton fluctuations lead to the adiabatic perturbations, while the curvaton carries the isocurvature (entropy) perturbations during inflation, and convert into the primordial density perturbations in the post-inflation era.

In this work, we consider the inflaton-curvaton mixed scenario, by assuming a slow-roll inflationary background where the inflaton potential is characterized by the slow-roll parameter $\epsilon \ll 1$. And $\epsilon$ is assumed to be time-independent under the quasi-de Sitter approximation. We generalize the SSR mechanism \cite{Cai:2018tuh} into the curvaton sector, i.e. the curvaton fluctuations propagate with a time-dependent oscillating sound speed during inflation. The modes around the characteristic wave lengths can be exponentially amplified due to the narrow resonance effect of SSR mechanism. Consequently, we acquire the enhanced primordial density perturbations on small scales, which is essential for PBH formation.

This article is organized as follows: in Section \ref{sec:SSRmix}, we briefly describe the inflaton-curvaton mixed scenario, and apply SSR mechanism in this scenario. The enhanced mode functions around the characteristic scales of the curvaton perturbations are obtained. After that, we calculate in detail the resonating inflaton and curvaton fluctuations under the de Sitter approximation and the quasi-de Sitter approximation, respectively. In Section \ref{sec:pertmix}, we discuss the generation of primordial density perturbations in our scenario, and numerically derive the constraints on the corresponding parameter space. The parametrized power spectrum is then obtained. In Section \ref{sec:PBHform}, we acquire the specific PBH mass spectrum which is within the current bounds of various experiments. We conclude by presenting our results with discussion in Section \ref{sec:concl}.

\section{SSR in the inflaton-curvaton mixed scenario}
\label{sec:SSRmix}

\subsection{Introduction to the inflaton-curvaton mixed scenario}

As mentioned previously, in the inflaton-curvaton mixed scenario, the total primordial curvature perturbation $\zeta$ on uniform-density hypersurfaces is contributed from both the inflaton perturbation and the curvaton perturbation \cite{Langlois:2004nn, Fonseca:2012cj}
\be \label{Pzeta_Mixed}
\mathcal{P}^\text{tot}_\zeta = \mathcal{P}_{\zeta}^\phi + \mathcal{P}_{\zeta}^\sigma ~,
\ee
where $\mathcal{P}_{\zeta}^\phi$ and $\mathcal{P}_{\zeta}^\sigma$ account for the contributions from inflaton $\phi$ part and curvaton $\sigma$ part, respectively. The primordial power spectrum $\mathcal{P}_{\zeta}^\phi$ is nearly scale-invariant on super-Hubble scales \cite{Mukhanov:1990me}
\be
\mathcal{P}_{\zeta}^\phi = \frac{1}{8 \pi^2 \epsilon } \l(\frac{H_*}{M_p}\r)^2 ~,
\ee
where $*$ refers to the quantity evaluated at Hubble-exit $k=aH$, and $M_p^2=1/8\pi G$ is the reduced Planck mass. In our work, the value of the slow-roll parameter is chosen as $\epsilon = 0.001$. Note that according to the observation \cite{Akrami:2018odb}, the scale-invariant power spectrum on large scales is measured as
\be \label{Constraint_Pzeta}
\mathcal{P}^\text{obs}_\zeta \sim 2 \times 10^{-9} ~,
\ee
which implies that $\mathcal{P}^\text{tot}_\zeta \simeq \mathcal{P}^\text{obs}_\zeta$ on large scales in the mixed case \eqref{Pzeta_Mixed}, and hence, the parameter choices ought to be in agreement with this requirement.

We further assume that the inflaton $\phi$ and the curvaton $\sigma$ are uncorrelated. 
For simplicity, we consider the case that the curvaton's potential is quadratic \footnote{ In a realistic model that leads to the oscillating sound speed of curvaton, this case would be changed when curvaton couples to inflaton or other field(s). These couplings in fact give a time-dependent effective mass $m_\text{eff}(t)$ to curvaton \cite{Lyth:2002my,Chen:2009we,Chen:2009zp}. Moreover, the curvaton's potential, in any case, virtually quadratic sufficiently close to the minimum \cite{Sasaki:2006kq}. And thus, our analyses in this paper are still valid when coupling exists.} , i.e.
\be \label{V}
V(\phi,\sigma) = V(\phi) + \frac12 m_\sigma^2 \sigma^2 ~,
\ee
where $V(\phi)$ is the inflaton potential and $m_\sigma$ is the mass of curvaton which is a positive constant. In the standard curvaton scenario, the curvaton is expected to be light ($m_\sigma \ll H$) and subdominant during inflation, while the inflaton drives the inflationary expansion of background. After inflation, the Hubble parameter drops into the regime $H \sim m_\sigma$, then the curvaton starts oscillating about the minimum of its potential. At this oscillating phase, the curvaton behaves like the pressure-less matter, such that the time-averaging energy density evolving as $\bar{\rho}_\sigma \propto a^{-3}$. Primordial density perturbations are generated from the curvaton fluctuations when curvaton decays. In the mixed case, these physical processes still hold. Note that, it is clear from the CMB observations that the primordial isocurvature perturbations had to be subdominant \cite{Enqvist:2000hp,Enqvist:2001zp}, while the adiabatic mode led to a good agreement with observations \cite{Ade:2015lrj,Akrami:2018odb}. Since the curvaton carries the entropy perturbations at the primordial epoch, making it possible to leave isocurvature perturbations after the curvaton decays. However, so long as the curvaton decays into radiation before primordial nucleosynthesis, the entropy perturbation can be converted to an adiabatic one. This also requires that all the species are in thermal equilibrium and that the baryon asymmetry is generated after the curvaton decays \cite{Lyth:2002my,Sasaki:2006kq}. To agree with the observational constraints on the isocurvature perturbations, we assume that the curvaton satisfies the above conditions to avoid any residual isocurvature perturbations.

\subsection{SSR mechanism applied in the inflaton-curvaton mixed scenario}

Inspired by the parametrization form suggested in \cite{Cai:2018tuh}, the sound speed square parameter for the curvaton field is time-evolving during inflation and is assumed to be parametrized as follows:
\be \label{cs}
c_s^2 = 1 -2 \xi [ 1-\cos(2k_*\tau) ], ~\text{with}~ \tau>\tau_i ~,
\ee
where $\xi$ is a small dimensionless quantity that measures the oscillation amplitude and $k_*$ is the oscillation frequency. Note that, $\xi < 1/4$ is required such that $c_s^2$ is positively definite, and the oscillation begins at $\tau_i$, where $k_*$ needs to be deep inside the Hubble radius with $|k_*\tau_i|\gg1$. Moreover, we set $c_s^2=1$ before $\tau_i$ and phenomenologically assume that it can begin to oscillate smoothly. And we also make an assumption that the sound speed square recovers $c_s^2=1$ smoothly at the end of inflation. The SSR mechanism was first studied in \cite{Cai:2018tuh}, in which the inflaton was assumed to undergo the SSR rather than the curvaton. Note that, the sound speed parameter $c_s$ of curvaton can deviate from unity during inflation via a non-canonical kinetic term. This may arise when curvaton models are embedded in some UV-complete theories, such as D-brane dynamics in string theory \cite{Silverstein:2003hf, Alishahiha:2004eh}, or k-essence \cite{ArmendarizPicon:1999rj, Garriga:1999vw}, or by introducing a coupling to another field from the effective field theory viewpoint when the heavy modes are integrated out \cite{Achucarro:2010da, Achucarro:2013cva, Pi:2017gih}. 
The study on the detailed realization is on-going in the follow-up work.

We start from the Mukhanov-Sasaki equation for the canonically normalized field perturbation \cite{Mukhanov:1988jd,Sasaki:1986hm}
\be \label{EoMv}
v_k'' + \l(c_s^2 k^2 - \frac{z''}{z} \r) v_k = 0 ~,
\ee
where the prime denotes the derivative with respect to the conformal time $\tau$ (i.e. $d\tau=dt/a$ and $a$ is the scale factor), and we have introduced the variables
\be \label{MS_Varible}
v \equiv \frac{a}{c_s} Q_\sigma, ~~z^2 = \frac{a^2}{c_s^2} \l( \frac{\bar{\sigma}'}{\mathcal{H}} \r)^2 ~,
\ee
where $\mathcal{H}$ is the comoving Hubble parameter and $Q_\sigma$ is the gauge-invariant curvaton perturbation
\be
Q_\sigma = \delta\sigma + \frac{\bar{\sigma}'}{\mathcal{H}} \psi ~.
\ee
Note that in the spatially flat gauge $\psi=0$, $Q_\sigma$ coincides with $\delta\sigma$.

For a homogeneous background, the curvaton satisfies the background equation of motion (EoM) as follows:
\be \label{EoM_sigmabar}
\chi'' + \l( a^2 m_\sigma^2 - \frac{a''}{a} \r) \chi = 0 ~,
\ee
where we have defined $\chi(\tau) \equiv a(\tau) \bar{\sigma}(\tau)$. In a quasi-de Sitter approximation, the scale factor behaves as $a(\tau) \simeq 1/ H (\epsilon - 1) \tau$, and the term $a''/a$ can be expanded in terms of the slow-roll parameter $\epsilon$, i.e. $a''/a = (2 + 3 \epsilon)/\tau^2 + \mathcal{O}(\epsilon^2)$ \cite{Bassett:2005xm}. Taking $\eta \equiv m_\sigma^2/H^2$ and expanding for small values of $\epsilon$ and $\eta$, EoM \eqref{EoM_sigmabar} can yield the solutions of a power-law form $\chi \propto \tau^{\alpha_{\pm}}$, where
\be \label{alpha}
\alpha_{\pm} = \frac12 \l(1 \pm \sqrt{-4 (-3 \epsilon + \eta) + 9} \r) ~,
\ee
where $\alpha_{-}$ and $\alpha_{+}$ denote a near-constant solution and a decreasing solution for $\bar{\sigma}$, respectively. 
According to the constraints \eqref{constraint_sigma} on the background curvaton field, the decreasing solution needs to be dropped. Then we have $\bar{\sigma}(\tau) \simeq \bar{\sigma}_* (\tau/\tau_*)^{1 + \alpha_-}$, where $\bar{\sigma}_*$ is evaluated at Hubble-exit of the relevant mode. Thus, the term $\bar{\sigma}'/\mathcal{H}$ is estimated as
\be
\begin{aligned}
	\frac{\bar{\sigma}'}{\mathcal{H}}
	\simeq \frac{\bar{\sigma}_*}{\tau_*^{1 + \alpha_{-}}} (\epsilon \alpha_{-} - \alpha_{-} + \epsilon -1) \tau^{1 + \alpha_{-}} ~.
\end{aligned}
\ee
By using the expression \eqref{alpha}, the effective mass term $z''/z$ can be expanded in terms of small amplitude $\xi$ and slow-roll parameter $\epsilon$
\be \label{z''/z}
\begin{aligned}
	\frac{z''}{z} =& \frac{\alpha_{-}^2 - \alpha_{-} +\epsilon - 2 \alpha_{-} \epsilon}{\tau^2}
	\\&+ \frac{4 k_* \xi}{\tau} (-\epsilon + \alpha_{-}) \sin(2 k_* \tau) + 4 k_*^2 \xi \cos(2 k_* \tau)
	+ \mathcal{O}(\xi^2,\epsilon^2) ~.
\end{aligned}
\ee

\begin{figure}[tbp]
	\centering
	\includegraphics[width=3in]{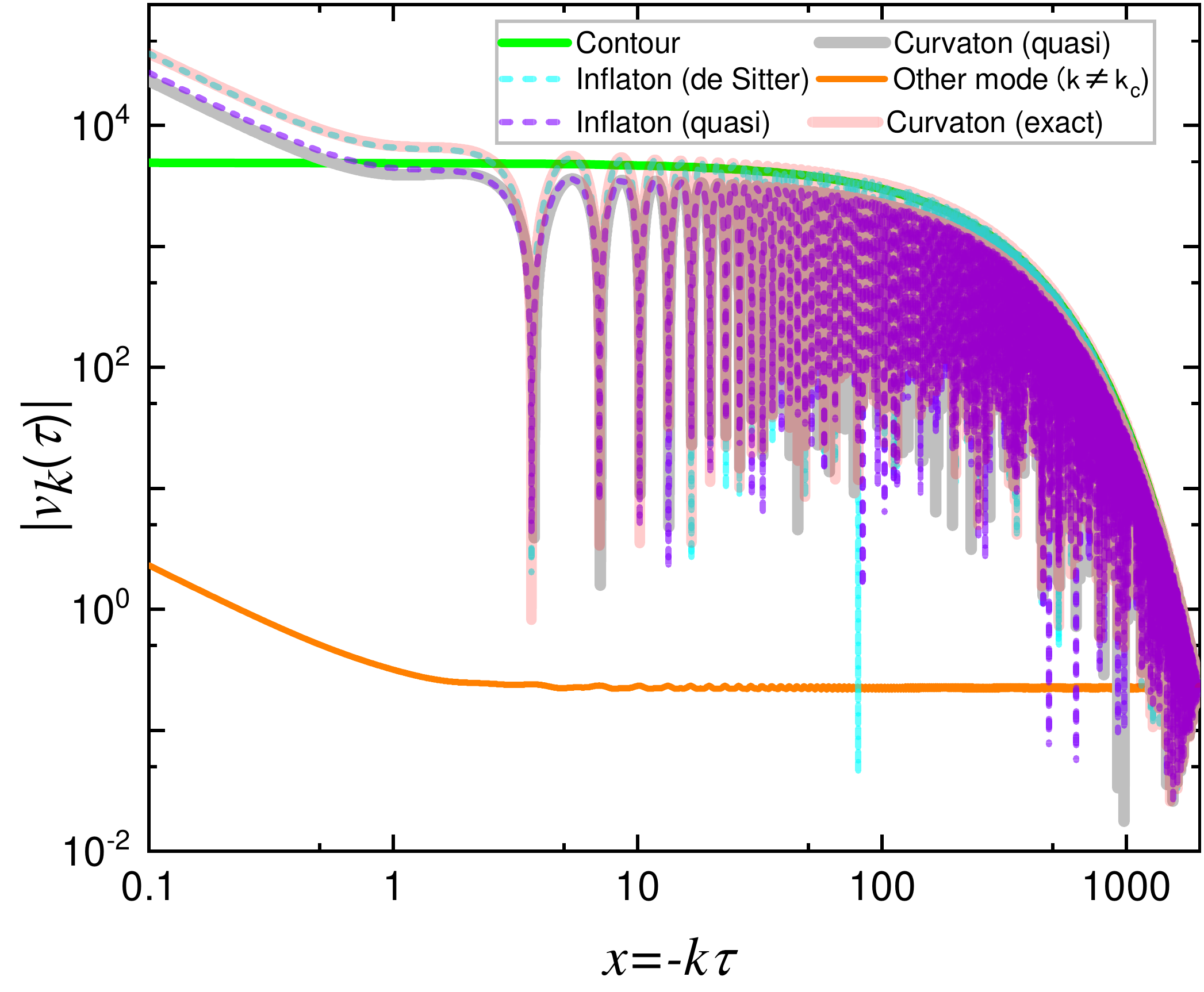}
	\caption{Comparison between the de Sitter approximation and the quasi-de Sitter approximation of resonating inflaton and curvaton whose modes are in the neighborhood of the characteristic scale $k_c$. Numerical results of the resonating inflaton in the de Sitter approximation and the quasi-de Sitter approximation are given by the cyan dashed line and the purple dashed line, respectively. One can see that these two results match well to a large extent. The numerical results of the resonating curvaton in the quasi-de Sitter approximation and under the Starobinsky inflation model are given by the grey solid line and the pink solid line, respectively. The orange solid line represents a mode $k \neq k_c$ that is not resonating. The growth of mode function is depicted by the green solid line, which can be described as $|v_{k_c}(\tau)| \propto \exp(\xi k_c \tau/2)$.}
	\label{fig_comparison}
\end{figure}

Note that in the de Sitter limit $\epsilon \rightarrow 0$ and using the assumption that the curvaton is very light during inflation, where $\eta \rightarrow 0$, the above expression of $z''/z$ \eqref{z''/z} reduces to the one in \cite{Cai:2018tuh}. In the SSR mechanism, the oscillation of sound speed square starts inside the Hubble radius with $|k\tau|\gg1$, and thus the first two terms in $z''/z$ \eqref{z''/z} become negligible on sub-Hubble scales. 
Consequently, Eq. \eqref{EoMv} can be approximated in the form of the Mathieu equation:
\be \label{Mathieu}
\frac{d^2 v_k}{dx^2} + (A_k - 2 q \cos2x) v_k = 0 ~,
\ee
where $x=- k_* \tau$, $A_k = \frac{k^2}{k_*^2} ( 1 -2 \xi )$ and $q = ( 2 - \frac{k^2}{k_*^2} ) \xi$, of which the solution is the combination of Mathieu cosine and sine functions, as shown explicitly in Appendix \ref{A1A2_app}. The solution can be exponentially amplified in some regions of parameter space, which is the key idea of enhancing the primordial fluctuations in SSR mechanism \cite{Cai:2018tuh}. Since the parameters $A_k$ and $q$ in the Mathieu equation \eqref{Mathieu} are the same with \cite{Cai:2018tuh}, both results of $v_k$ should coincide with each other on sub-Hubble scales. 
The solution of Eq. \eqref{Mathieu} does not manifestly rely on the inflationary background on sub-Hubble scales since this equation does not involve any background model parameters such as the slow-roll parameter $\epsilon$. On super-Hubble scales, our numerical results show that the resonating inflaton and resonating curvaton match reasonably well, as shown in Fig. \ref{fig_comparison}. 

We mention that around the characteristic scale $k_c \simeq (1+\xi) k_*$, the mode function $v_{k_c}(\tau)$ oscillates and increases inside the Hubble radius, while on super-Hubble scales, $v_{k_c}(\tau)$ behaves as $\sim 1/\tau$, and thus the curvaton field fluctuation $\delta\sigma$ is nearly frozen. For other modes $k \neq k_c$, the mode function is not amplified, which behaves like the Bunch-Davis (BD) vacuum \cite{Cai:2018tuh}. Note that in the Mathieu equation \eqref{Mathieu}, $\xi$ is small and thus $|q|\ll 1$, implying that resonance bands are located in narrow ranges about the harmonic frequencies $k \simeq n k_c$ with $n$ being an arbitrary integer number. This is the narrow resonance effect for the SSR mechanism. The growth of $v_{k_c}(\tau)$ can be estimated by
\be \label{ModeFunction}
|v_{k_c}(\tau)| \propto \exp(\xi k_c \tau/2) ~,
\ee
which is denoted by a green solid curve in Fig. \ref{fig_comparison}. The amplification ceases around Hubble-exit, since the first term in Eq. \eqref{z''/z} becomes dominant on super-Hubble scales.

The above analyses show that, in the SSR mechanism, the behaviours of mode functions $v_{k}(\tau)$ of resonating inflaton and curvaton are very similar under the quasi-de Sitter approximation, which are presented in Fig. \ref{fig_comparison}. We also find that the de Sitter approximation and the quasi-de Sitter approximation match well for resonating inflaton in the previous work \cite{Cai:2018tuh}. 
For the exact solution of resonating curvaton in a quasi-de Sitter background, one needs to choose a specific model of inflation, and numerically solve the Mukhanov-Sasaki equation \eqref{EoMv} combining with the background equation \eqref{EoM_sigmabar}. For instance, we present the numerical result for the Starobinsky inflation model \cite{Starobinsky:1980te} which is represented by the pink solid curve in Fig. \ref{fig_comparison}. The numerical solution of $v_k(\tau)$ in this model matches the solution in the quasi-de Sitter approximation (depicted by the grey solid line) on both sub-Hubble and super-Hubble scales, as shown in Fig. \ref{fig_comparison}.

Note that the amplified mode function \eqref{ModeFunction} will lead to enhanced curvature perturbation after inflation through the conversion of entropy perturbations carried by curvaton. In the following section, we will discuss this aspect in detail.

\section{Primordial density perturbations in the inflaton-curvaton mixed scenario}
\label{sec:pertmix}

\subsection{Primordial curvature perturbation from resonating curvaton}

On super-Hubble scales, the curvature perturbations on uniform-density hypersurfaces can be written as \cite{Lyth:2004gb,Sasaki:2006kq}
\be \label{zeta}
\begin{aligned}
	\zeta_i(t,\mathbf{x}) 
	= \delta N(t,\mathbf{x}) + \frac13 \int_{\bar{\rho}_i(t)}^{\rho_i(t,\mathbf{x})} \frac{d\tilde{\rho}_i}{\tilde{\rho}_i + P_i(\tilde{\rho}_i)} ~,
\end{aligned}
\ee
where the subscript $i$ denotes either inflaton $\phi$, curvaton $\sigma$ or total energy density. $\omega_i = P_i/\rho_i$ is the equation-of-state parameter, and $\rho_i$ ($\bar{\rho}_i$) denotes the (background) energy density of the $i$-th field component. Using Eq. \eqref{zeta}, the local density of $i$ can be written in terms of its homogeneous value and the inhomogeneous expansion perturbation $\delta N$
\be \label{rho_i}
\rho_i = \bar{\rho}_i e^{3(1+\omega_i) (\zeta_i - \delta N)} ~.
\ee

During the early radiation-dominated phase, the curvaton oscillates around the minimum of its potential, but before it decays, it behaves like the pressureless matter ($\bar{\rho}_\sigma \propto a^{-3}$), and then its time-averaged energy density is \cite{Langlois:2008vk,Lyth:2002my}
\be \label{rho_s}
\rho_\sigma = m_\sigma^2 \sigma^2 ~,
\ee
where $\sigma$ is the root-mean-square amplitude of the curvaton field. Using the equation \eqref{rho_i} for the oscillating curvaton, we have
\be
\rho_\sigma = \bar{\rho}_\sigma e^{3 (\zeta_\sigma - \delta N)} ~.
\ee
In the post-inflation era, the curvaton is still subdominant, the spatially flat hypersurfaces are characterized by $\delta N=\zeta_\phi$ \cite{Langlois:2008vk}, where $\zeta_\phi$ corresponds to the adiabatic perturbation generated by the inflaton fluctuations $\delta\phi$, then we yield
\be \label{rho_s_e}
\rho_\sigma = \bar{\rho}_\sigma e^{3 (\zeta_\sigma - \zeta_\phi)} ~.
\ee
The entropy perturbation $\mathcal{S}_\sigma$ carried by curvaton perturbation is given by $\mathcal{S}_\sigma \equiv 3 (\zeta_\sigma - \zeta_\phi)$. Combining the relations \eqref{rho_s} and \eqref{rho_s_e}, one yields
\be
m_\sigma^2 (\bar{\sigma} + \delta\sigma)^2 = m_\sigma^2 \bar{\sigma}^2 e^{\mathcal{S}_\sigma} ~.
\ee
At the linear order, one acquires
\be \label{entropy}
\mathcal{S}_\sigma = 2 \frac{\delta\sigma}{\bar{\sigma}} ~,
\ee
which is consistent with \cite{Sasaki:2006kq,Cai:2010rt} in which $\zeta_\phi=0$. 

In the inflaton-curvaton mixed scenario, the sound speed is assumed to transit to $c_s^2=1$ smoothly at the end of inflation. And we also assume that the curvaton is subdominant and weakly coupled before it decays, the dynamics of $\delta\sigma$ can be written as
\be \label{eom_delta_sigma}
\delta\sigma'' + 2 \mathcal{H} \delta\sigma' + a^2 m_\sigma^2 \delta\sigma = 0 ~.
\ee
Note that comparing with \eqref{EoM_sigmabar}, $\bar{\sigma}$ and $\delta\sigma$ share the same EoM. Hence, the fraction $\delta\sigma/\bar{\sigma}$ remains constant on super-Hubble scales before the curvaton decays \cite{Lyth:2001nq,Lyth:2002my,Kohri:2012yw}. Thus, we evaluate this fraction at the end of inflation, and the entropy perturbation \eqref{entropy} is rewritten as
\be \label{entropy_initial}
\mathcal{S}_\sigma = 2 \frac{\delta\sigma_\text{end}}{\bar{\sigma}_\text{end}} \simeq 2 \frac{\delta\sigma_*}{\bar{\sigma}_*} ~.
\ee
Since $\delta\sigma$ and $\bar{\sigma}$ are frozen on super-Hubble scales during inflation. In the mixed case, we regard $\bar{\sigma}_*$ and $\delta\sigma_*$ as the initial conditions \footnote{ Note that, if one considers an non-canonical kinematic term or a coupling to other field(s) for a curvaton model that leads to the resonance effect, $\bar{\sigma}$ and $\delta\sigma$ may not share the same EoM like equations \eqref{EoM_sigmabar} and \eqref{eom_delta_sigma}. In this case, the entropy \eqref{entropy_initial} carried by curvaton will becomes $S_\sigma \simeq 2 q \frac{\delta\sigma_*}{\bar{\sigma}_*} $, where the time-dependent parameter $q$ accounts for the evolution of the fraction $\frac{\delta\sigma}{\bar{\sigma}}$ from the Hubble exit to the curvaton decay \cite{Lyth:2002my,Sasaki:2006kq}. One needs to resort to a numerical analysis for the dynamics of $\delta\sigma$ and $\bar{\sigma}$ to derive the c-number $q(t_\text{dec})$ when the curvaton decays \cite{Langlois:2004nn}. And thus, $q(t_\text{dec})$ does not deform the shape (i.e., the narrow resonance) of the resulting amplified primordial density power spectrum. Moreover, as we will discuss later, the amplitude of primordial density power spectrum in SSR mechanism is largely depend on the amplitude of sound speed $\xi$ and evolution of the Universe, the effect of $q(t_\text{dec})$ on the efficiency of the resonance effect in SSR mechanism can be safely neglected. }.

One of constraints on $\bar{\sigma}_*$ comes from the expectation that the quantum fluctuations in a weakly coupled field such as the curvaton $\delta\sigma_*$, to be well described by a Gaussian random field \cite{Sasaki:2006kq}. Then $\bar{\sigma}_*$ satisfies \cite{Lyth:2001nq}
\be
H_* \ll \bar{\sigma}_* ~.
\ee
In addition, $\sigma$ is supposed to start oscillating before it decays and still strongly subdominant with respect to radiation (in order to avoid the secondary inflation triggered by curvaton \cite{Langlois:2004nn,Dimopoulos:2011gb}). This implies \cite{Langlois:2004nn}
\be
\bar{\sigma}_* \ll M_p ~.
\ee
Thus, we now have the constraints on $\bar{\sigma}_*$
\be \label{constraint_sigma}
H_* \ll \bar{\sigma}_* \ll M_p ~.
\ee

We assume that the inflaton completely decays into radiation immediately at the end of inflation. Any initial inflaton perturbation would lead to a perturbation in radiation energy density perturbation before the curvaton decays, we denote it by $\rho_R$ \cite{Langlois:2008vk}. Using the equation \eqref{rho_i}, we write
\be
\rho_\sigma = \bar{\rho}_\sigma e^{3 (\zeta_\sigma - \delta N)},~~
\rho_R = \bar{\rho}_R e^{4 (\zeta_\phi - \delta N)} ~.
\ee

For simplicity, we also assume the sudden-decay approximation for curvaton, i.e., the curvaton decays on a uniform-total density hypersurface when $H=\Gamma_\sigma$, where $\Gamma_\sigma$ is the decay rate of the curvaton (assumed to be constant). The results presented in \cite{Sasaki:2006kq} show that the sudden-decay approximation matches well with the gradual decay approximation. The calculation could be performed on other hypersurfaces, e.g. a uniform curvaton density hypersurface \cite{Cai:2010rt}. Thus, on this uniform-total density hypersurface, we have
\be \label{EnergyConservation}
\rho_R(t_\text{dec},\mathbf{x}) + \rho_\sigma(t_\text{dec},\mathbf{x}) = \bar{\rho}_r(t) ~,
\ee
where $\bar{\rho}_r(t)$ is the total background energy density when the curvaton decays. Note that $\delta N = \zeta_r$ according to the equation \eqref{rho_i}, and $\zeta_r$ is the total curvature perturbation at decay of curvaton. Assuming all the curvaton decay products are relativistic, then $\zeta_r$ is conserved after the curvaton decay and before Hubble re-enter. The equation \eqref{EnergyConservation} gives \cite{Sasaki:2006kq}
\be \label{Omega_sigma}
\Omega_{\sigma,\text{dec}} e^{3(\zeta_\sigma - \zeta_r)} + (1 - \Omega_{\sigma,\text{dec}}) e^{4(\zeta_\phi - \zeta_r)} = 1 ~,
\ee
where $\Omega_{\sigma,\text{dec}} = \bar{\rho}_\sigma/(\bar{\rho}_\sigma + \bar{\rho}_R)$ is the curvaton density fraction at its decay. The equation \eqref{Omega_sigma} gives a fully nonlinear relation between the primordial curvature perturbation $\zeta_r$, the inflaton perturbation $\zeta_\phi$ and the curvaton perturbation $\zeta_\sigma$. At the linear order, \eqref{Omega_sigma} gives
\be \label{zeta_r}
\zeta_r = f \zeta_\sigma + (1 - f) \zeta_\phi = \zeta_\phi + \frac{f}{3} \mathcal{S}_\sigma ~,
\ee
where the energy fraction $f$ is defined as
\be \label{fraction_f}
f = \frac{3 \Omega_{\sigma,\text{dec}}}{4 - \Omega_{\sigma,\text{dec}}} = \frac{3 \bar{\rho}_\sigma}{3 \bar{\rho}_\sigma + 4 \bar{\rho}_R} ~.
\ee

Using \eqref{entropy} and \eqref{zeta_r}, the total power spectrum for the primordial curvature perturbation is written as
\be \label{Pzeta_r}
\begin{aligned}
	\mathcal{P}_{\zeta_r}
	=& \mathcal{P}_{\zeta_\phi} + \frac{f^2}{9} \mathcal{P}_{\mathcal{S}_\sigma}
	\\=& \frac{1}{8 \pi^2 \epsilon } \l(\frac{H_*}{M_p}\r)^2 
	+ \frac{2}{9\pi^2} f^2 (\epsilon-1)^2 c_s^2(\tau_*) k | v_k(\tau_*) |^2 \l( \frac{H_*}{\bar{\sigma}_*} \r)^2 ~,
\end{aligned}
\ee
where $\tau_* = -1/k$. It is useful to introduce dimensionless parameter to measure the relative contribution of curvaton to that of inflaton \cite{Langlois:2008vk}
\be
\lambda 
= \frac{16}{9} f^2 \epsilon (\epsilon-1)^2 c_s^2(\tau_*) k \l| v_k(\tau_*) \r|^2 \l( \frac{M_p}{\bar{\sigma}_*} \r)^2 ~,
\ee
then the total primordial power spectrum is given by
\be \label{Ptotal}
\mathcal{P}^\text{tot}_\zeta = \mathcal{P}_{\zeta_r} = (1 + \lambda) \mathcal{P}_{\zeta_\phi} ~.
\ee

\subsection{Constraints on model parameter space}

\begin{figure}[tbp]
	\centering
	\includegraphics[width=2.8in]{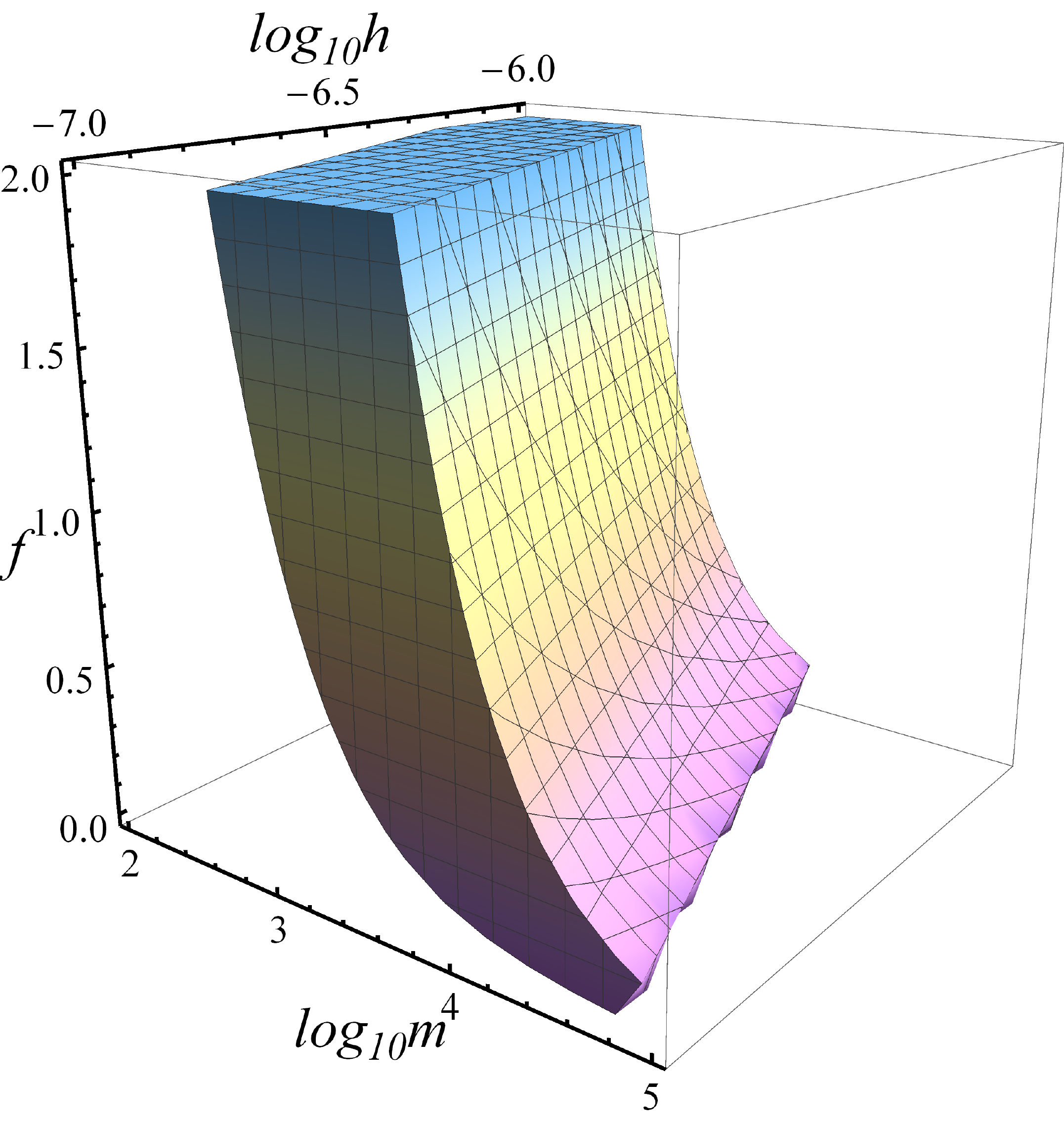}
	\caption{Observational constraints on the parameter space $\{f,m,h\}$ of the inflaton-curvaton mixed scenario. The parameter space is constrained by the scale-invariant power spectrum \eqref{Constraint_Pzeta} $\mathcal{P}^\text{obs}_\zeta \simeq 2 \times 10^{-9}$ on large scales, while the major peak of the enhanced power spectrum $\mathcal{P}^\text{tot}_\zeta(k_c)$ \eqref{Ptotal} on small scales is bounded by 1 in the perturbative regime.}
	\label{fig_constaint}
\end{figure}

The constraints on the mixed scenario arise from two aspects: on large scales, the observational constraints \eqref{Constraint_Pzeta} imply that $\mathcal{P}^\text{tot}_\zeta \simeq \mathcal{P}^\text{obs}_\zeta \simeq 2 \times 10^{-9}$; And on small scales, the power spectrum is exponentially amplified around the characteristic scale $k_c$ due to the narrow resonance effect of SSR mechanism. As we work in the perturbative regime, the magnitude of total primordial power spectrum should be limited by unity, i.e. $ \mathcal{P}^\text{tot}_\zeta(k_c) < 1 $. The constraints on the parameter space of the mixed scenario are shown in Fig. \ref{fig_constaint}. Besides the parameters in SSR mechanism $\{\xi,\tau_i,k_*\}$ \cite{Cai:2018tuh}, the set of parameter space is given by $\{f,m,h\}$, where the energy fraction $f$ \eqref{fraction_f} is the energy fraction of curvaton field; while $m \equiv M_p/\bar{\sigma}_* $ and $h \equiv H_*/M_p$. Taking into account of the constraints from observations \eqref{Constraint_Pzeta} and that of curvaton field \eqref{constraint_sigma}, we find such constraints of $h < 1.3 \times 10^{-5}$ and $1 \ll m \ll 10^5$.

Note that, the above two constraints arise from the linear level, i.e., we study the fully nonlinear relation \eqref{Omega_sigma} at the linear level in the previous discussion, while the higher-order terms give rise to the non-Gaussianity of the full curvature perturbation $\zeta$ in curvaton scenario \cite{Sasaki:2006kq}. The non-Gaussianity of primordial perturbation is sensitively bounded by the current CMB observations \cite{Akrami:2019izv}, this shall also provide a constraint on our mixed model. Many previous works (e.g., see \cite{Lyth:2002my,Langlois:2004nn,Sasaki:2006kq,Fonseca:2012cj}) show that the large non-Gaussianity could be generated in curvaton scenario or inflaton-curvaton mixed scenario, if the curvaton contributes only a small fraction of the energy density before it decays, i.e., $f \ll 1$. Thus, it is expected that non-Gaussianity could constrain the range of the energy density fraction $f$ defined in \eqref{fraction_f}. The non-Gaussianity is usually measured by the local nonlinear parameters $f_\text{NL}$ and $g_\text{NL}$ defined by $\zeta = \zeta_G + (3/5) f_\text{NL} \zeta_G^2 + (9/25) g_\text{NL} \zeta_G^3$, where $\zeta_G$ represents the Gaussian primordial perturbation. {\it Planck 2018} experiment gives the constraints \cite{Akrami:2019izv}:
\be \label{bound_NG}
f_\text{NL} = -0.9 \pm 5.1 (68\% ~ \text{CL}),
~~
g_\text{NL} = (-5.8 \pm 6.5) \times 10^4 (68\% ~ \text{CL}) ~.
\ee
In our mixed scenario, the sound speed of curvaton stops oscillating at the end of inflation, and the physical processes of curvaton after inflation are same with the previous inflaton-curvaton mixed scenario discussed in \cite{Langlois:2004nn,Fonseca:2012cj}. So that, the formulas for non-Gaussianity derived there are still hold in our mixed scenario. The local nonlinear parameters $f_\text{NL}$ and $g_\text{NL}$ are given by \footnote{For a non-quadratic potential, one needs to consider the nonlinear evolution of curvaton field perturbation between Hubble exit and the oscillation of curvaton \cite{Sasaki:2006kq,Fonseca:2012cj}.}
\be \label{fNL}
f_\text{NL} = \l(\frac{5}{4 f} - \frac{5}{3} - \frac{5}{6} f \r) \l(\frac{\lambda}{1 + \lambda} \r)^2 ~,
\ee
and 
\be \label{gNL}
g_\text{NL} = \frac{25}{54} \l( - \frac{9}{f} + \frac12 + 10 f + 3 f^2 \r) \l(\frac{\lambda}{1 + \lambda} \r)^3 ~,
\ee
where we have assumed that the energy density $f$ is independent of curvaton field $\bar{\sigma}_*$ for the quadratic curvaton potential \eqref{V}. It should be noted that there are $1/f^2$ terms in $f_\text{NL}$ and $g_\text{NL}$, making it possible to produce large non-Gaussianity in inflaton-curvaton mixed scenario \cite{Lyth:2002my,Langlois:2004nn,Sasaki:2006kq,Fonseca:2012cj}.

For the non-resonant region ($k \neq k_c$), \eqref{fNL} and \eqref{gNL} reads off
\be
f_\text{NL} = \l(\frac{5}{4 f} - \frac{5}{3} - \frac{5}{6} f \r) \l[ \frac{\frac89 \epsilon (\epsilon - 1)^2 (f m)^2}{1 + \frac89 \epsilon (\epsilon - 1)^2 (f m)^2} \r]^2 ~,
\ee
and
\be
g_\text{NL} = \frac{25}{54} \l( - \frac{9}{f} + \frac12 + 10 f + 3 f^2 \r) \l[ \frac{\frac89 \epsilon (\epsilon - 1)^2 (f m)^2}{1 + \frac89 \epsilon (\epsilon - 1)^2 (f m)^2} \r]^3 ~.
\ee
While for the resonant region ($k = k_c$), $\lambda$ is exponential amplified due to the narrow resonance of SSR mechanism, and then \eqref{fNL} and \eqref{gNL} are approximated to
\be
f_\text{NL} \simeq \frac{5}{4 f} - \frac{5}{3} - \frac{5}{6} f ~,
\ee
and
\be
g_\text{NL} \simeq \frac{25}{54} \l( - \frac{9}{f} + \frac12 + 10 f + 3 f^2 \r) ~,
\ee
which recover the standard curvaton results \cite{Sasaki:2006kq}. For $m= 3.4 \times 10^4$, the observational bounds \eqref{bound_NG} provide the constraints on $f$:
\be \label{f_range}
f \in \l\{
\begin{aligned}
	&(0, 0.00018) \cup (0.207, 5.474), ~~~ k \neq k_c,
	\\& (0.207, 5.474), ~~~ k = k_c ~~.
\end{aligned} \r.
\ee
Note that the range of value $(0.207, 5.474)$ is independent of the value of $m$. And we have assumed that the observational bounds \eqref{bound_NG} are also hold on small scales, which is not necessarily the true case. However, this implies that the non-Gaussianity can be used as a tool to test our mixed scenario in future experiments. For example, for the very small energy fraction $f = 0.04$, and $f_\text{NL} = 29.55$ and $g_\text{NL} = -103.505$ in the resonant region. With above considerations, we choose the range $(0, 0.00018) \cup (0.207, 5.474)$ of $f$ within the observational bounds.

\begin{figure}[tbp]
	\centering
	\includegraphics[width=3.2in]{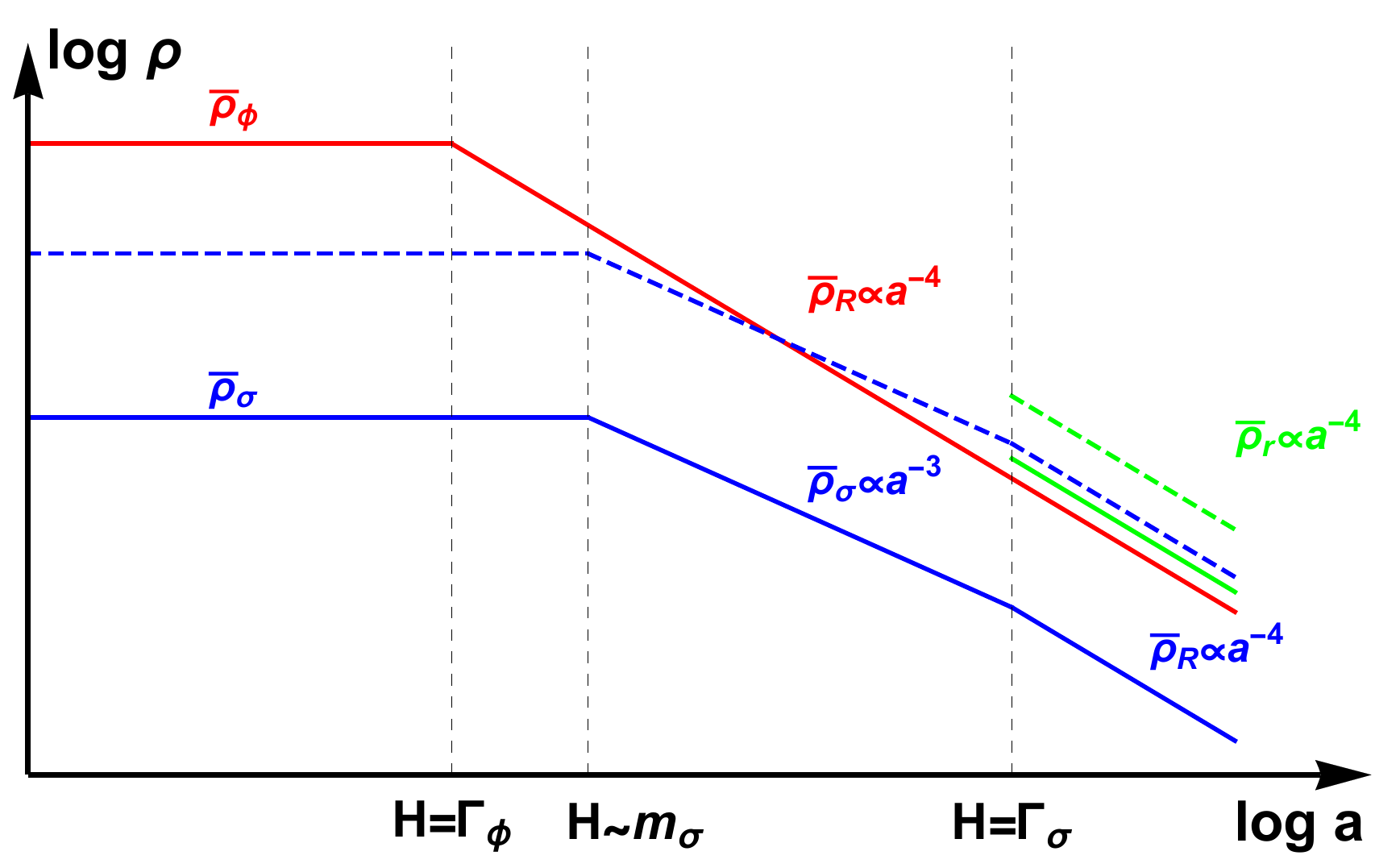}
	\caption{The diagram shows the evolution of the background energy density for inflaton (red line), curvaton (blue line), and their decay products (radiation). The green line denotes the total energy density of radiation after {\bf curvaton decays}. The solid lines refer to the case where curvaton is still subdominant at its decay, while the dashed lines refer to the case where curvaton becomes dominant before its decay.}
	\label{fig_evolution}
\end{figure}

\begin{figure}[tbp]
	\centering
	\includegraphics[height=2.6in]{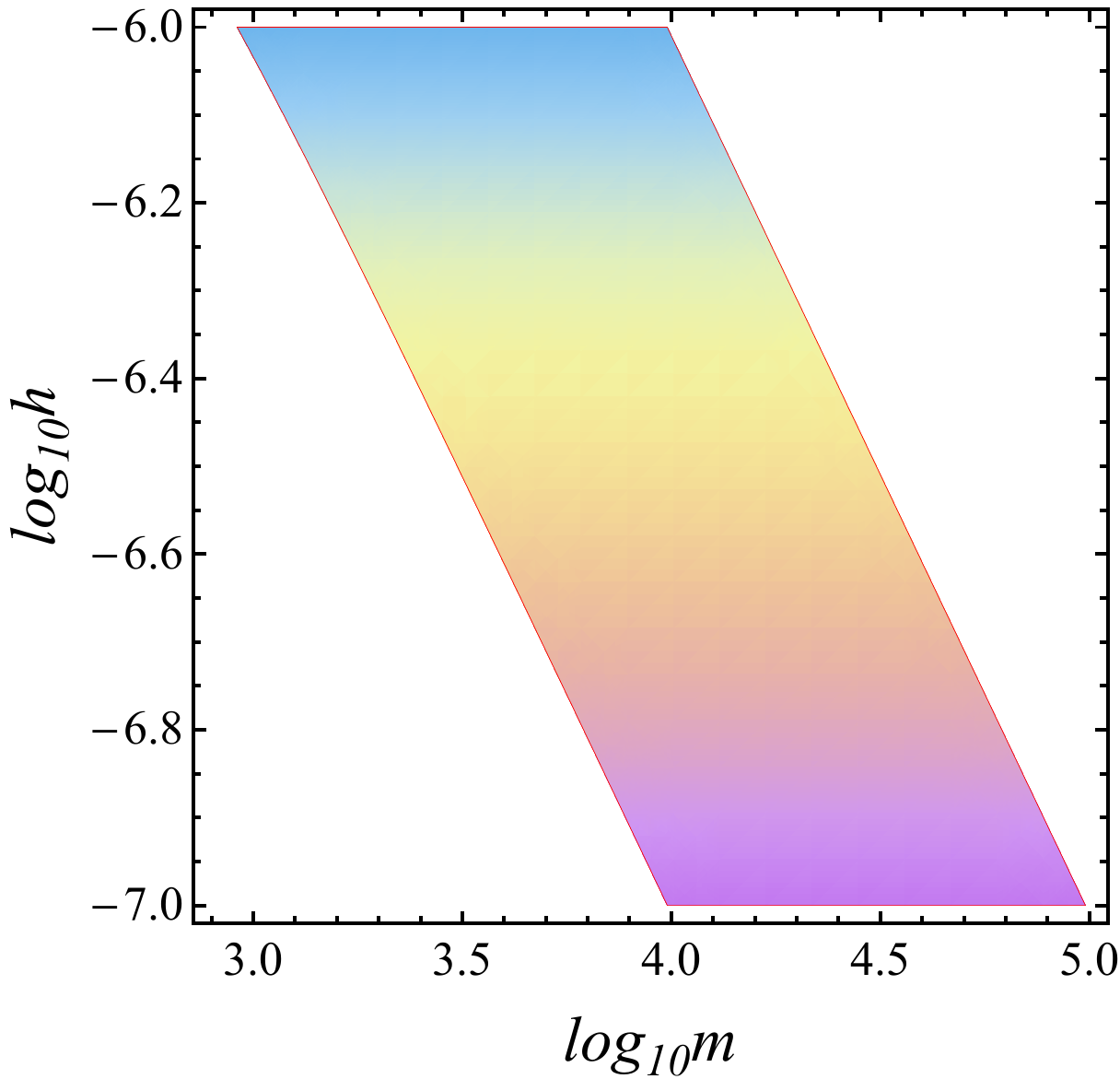}
	\caption{The constraints on the parameters $\{m,h\}$, when we take the values of parameters as follows: $\xi=0.1, \epsilon=0.001, v=200, f=0.04$.}
	\label{fig_0.04}
\end{figure}

\begin{figure}[tbp]
	\centering
	\includegraphics[width=3.3in]{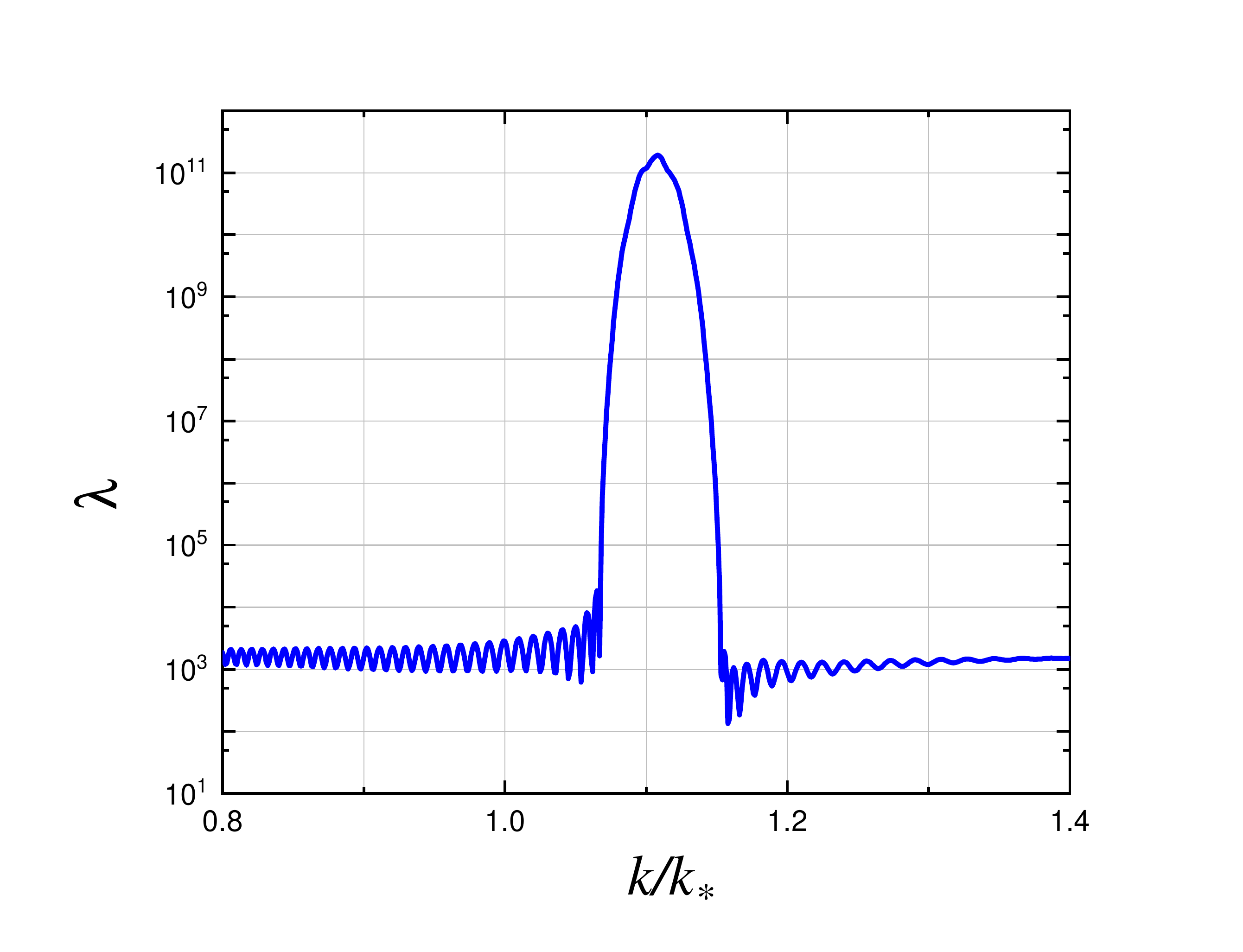}
	\caption{The relative fraction of contribution from curvaton to that of inflaton. The resonance happens around the characteristic mode $k_c \simeq (1 + \xi) k_*$, and the width of resonance is quite narrow ($\sim \xi k_*$). Within a wide range, we have $\lambda \gg 1$ which implies that the contribution of curvaton is dominant over inflaton for small energy fraction $f$. The parameters are chosen as follows: $\xi=0.1$, $\epsilon = 0.001$, $v= 200$, $\eta = 10^{-9}$, $f = 0.04$, $m= 3 \times 10^4$ and $h = 3.5 \times 10^{-7}$.}
	\label{fig_lambda}
\end{figure}

Fig. \ref{fig_evolution} shows the schematic diagram on the energy density evolutions of inflaton, curvaton and their decay products (radiation). Fig. \ref{fig_0.04} shows the parameter space of $\{m,h\}$ by setting $f=0.04$. Consequently, our numerical results show that $\lambda\gg 1$, could match the scale-invariant power spectrum \eqref{Constraint_Pzeta} on large scales, see Fig. \ref{fig_lambda}. This implies that the inflaton perturbation is subdominant for small energy fraction $f$ in the inflaton-curvaton mixed scenario \eqref{Pzeta_Mixed}, which recovers the standard curvaton models. 
Previous works on PBH formation via curvaton scenario \cite{Yokoyama:1995ex,Kawasaki:2012wr} suggest that in order to maintain the scale-invariant property, the primordial density perturbations generated from curvaton should be dominant on small scales while inflaton dominates on large scales. By contrast, our model shows that such perturbation could be dominant on all scales. For the large energy fraction $f$, $\lambda$ could be comparable with or even smaller than unity in the non-resonant region ($k \neq k_c$). For example, when we choose the values of parameters as follows: for $f = 1.5$, $m = 25$, $h = 9 \times 10^{-6}$, we yield $\lambda = \frac89 \epsilon (\epsilon - 1)^2 (f m)^2 \simeq 1.2$, implying the contributions from curvaton and of inflaton are on the same order of magnitude in the non-resonant region; for $f=2$, $m=10$, $h= 1.1 \times 10^{-5}$, we yield $\lambda \simeq 0.3$, implying that the relative contribution from inflaton is dominant on all scales except for the resonant region around $k_c$. Within the model parameter space $\{f, m, h\}$, the peak of $\lambda$ for the large energy fraction $f$ is slightly lower than that of the small $f$.

\subsection{Primordial density perturbation with a peak}

\begin{figure}[tbp]
	\centering
	\includegraphics[width=3.5in]{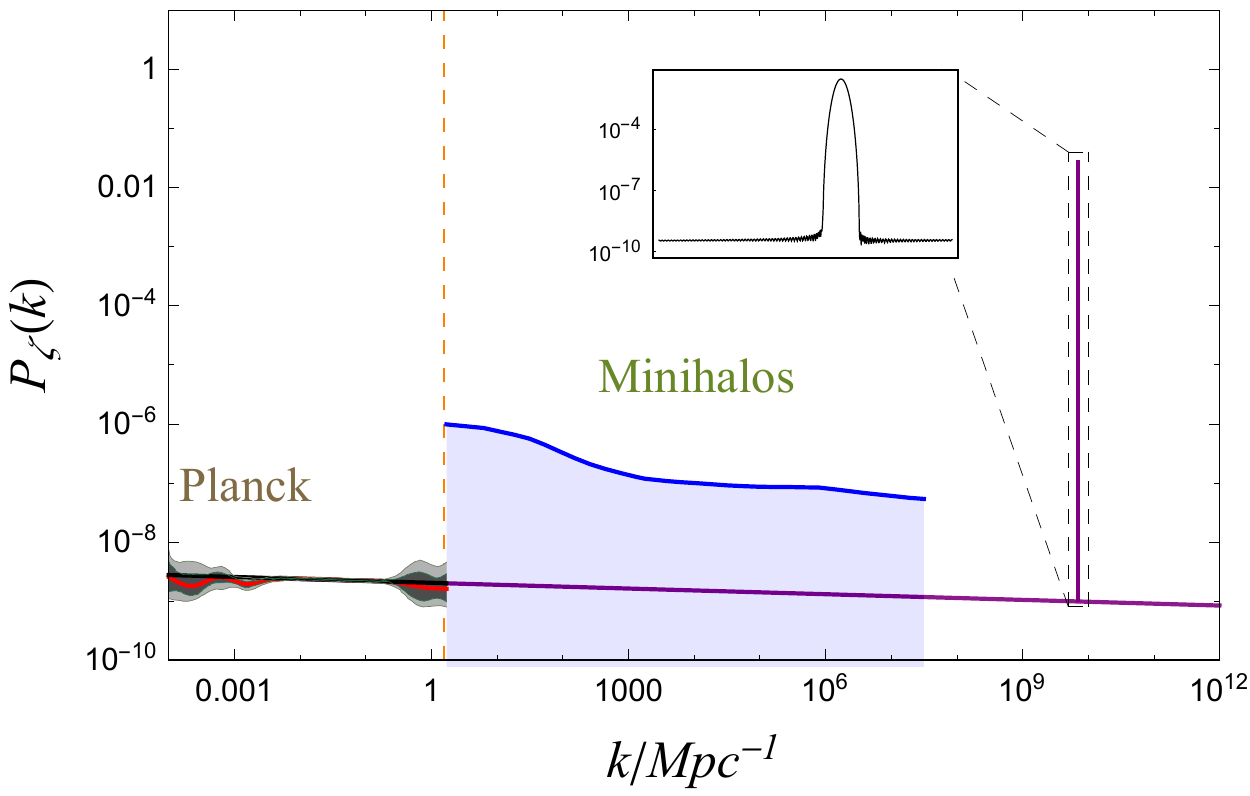}
	\caption{The power spectrum of primordial curvature perturbations with a sharp peak caused by SSR mechanism in the  inflaton-curvaton mixed scenario and the comparison with various observational windows \cite{Bringmann:2011ut}. The major peak is around $k_* = 7 \times 10^9$ Mpc $^{-1}$. The rest of the peaks are suppressed by a factor of $\mathcal{O}(10^{-8})$, which are not shown in the figure.}
	\label{fig_Pzeta}
\end{figure}

We focus on the characteristic mode $k_c \simeq (1 + \xi) k_*$, the parameter $\lambda$ evolves as $\lambda(\tau) \simeq \lambda(\tau_i) e^{-\xi k_c \tau_i} \simeq \frac89 \epsilon (\epsilon - 1)^2 (f m)^2 e^{-\xi k_c \tau_i}$ before Hubble crossing, where we have set the mode function at the beginning of resonance to the BD vacuum, $v_k(\tau_i) = e^{- i k \tau_i} / \sqrt{2 k}$. The enhancement factor $e^{-\xi k_c \tau_i}$ arises from the interplay of two effects as discussed in \cite{Cai:2018tuh}: the oscillation in the sound speed, controlled by its amplitude $\xi$; and the expansion of the Universe from the beginning of the resonance to the Hubble crossing of $k_c$, i.e., $ v \equiv -k_c \tau_i = \tau_i / \tau_c \simeq e^{\Delta N}$, where $\Delta N$ is the e-folding number for this period of inflation. The numerical results show that the peaks for harmonic frequencies $2k_c, 3k_c, \cdots$ are suppressed by a factor of around $\mathcal{O}(10^{-8})$, so they can be safely neglected. For simplicity, we keep our discussion on the $k_c$ mode, and parametrize the power spectrum using a $\delta$ function
\be \label{Pzeta_total}
\mathcal{P}_{\zeta}^\text{tot} \simeq A_s \l(\frac{k}{k_p}\r)^{n_s-1} \l( 1 + \frac{\xi k_*}{2} e^{ \xi v} \delta(k - k_c) \r) ~,
\ee
where $A_s = (1 + \lambda(\tau_i)) \mathcal{P}_{\zeta_\phi} \simeq 2 \times 10^{-9}$ is the amplitude of the standard power spectrum \eqref{Constraint_Pzeta} and $n_s$ is the spectral index at pivot scale $k_p \simeq 0.05$Mpc$^{-1}$. The total primordial power spectrum $\mathcal{P}_{\zeta}^\text{tot}$ is shown in Fig. \ref{fig_Pzeta}, the value is amplified around the characteristic scale $k_c$, while it remains nearly scale-invariant on large scales.

\section{PBH Formation}
\label{sec:PBHform}

We now study the formation of PBHs due to enhancement in the primordial power spectrum in the inflaton-curvaton mixed scenario \eqref{Pzeta_total}, the analysis is quite similar with \cite{Cai:2018tuh}. From Fig. \ref{fig_lambda}, the peak in the power spectrum is extremely narrow ($\sim \xi k_*$), thus only the modes with frequencies very close to $k_c$ would be able to reach sufficiently large amplitude to collapse into black holes. After Hubble exit, if the density perturbations from these modes are greater than a critical value $\delta_c$, they could collapse into black holes due to gravitational attraction after Hubble reentry. The Schwarzschild radius of PBHs with mass $M$ is related to the physical wavelength of the mode $k_M$ at a Hubble reentry, $k_{M,\text{ph}} = k_M / a_M \simeq R_s^{-1} = (M / 4 \pi M_p^2)^{-1}$. Accordingly, the PBH mass can be expressed as a function of $k_M$ via
\be
M \simeq \gamma \frac{4 \pi M_p^2}{H[t_\text{exit}(k_M)]} e^{\Delta N(k_M)} ~,
\ee
where $\Delta N(k_M) = \ln \{ a[t_\text{re-entry}(k_M)] / a[t_\text{exit}(k_M)] \}$ is the e-folding number from the Hubble-exit time of the mode $k_M$ to its reentry time. The correction factor $\gamma$ represents the fraction of the horizon mass responsible for PBH formation, which can be simply taken as $\gamma \simeq 0.2$ \cite{Carr:1975qj}. Given the sharpness of the peak in the power spectrum, the PBHs formed in this context are likely to possess a rather narrow range of mass. 

\begin{figure}[tbp]
	\centering
	\includegraphics[width=3in]{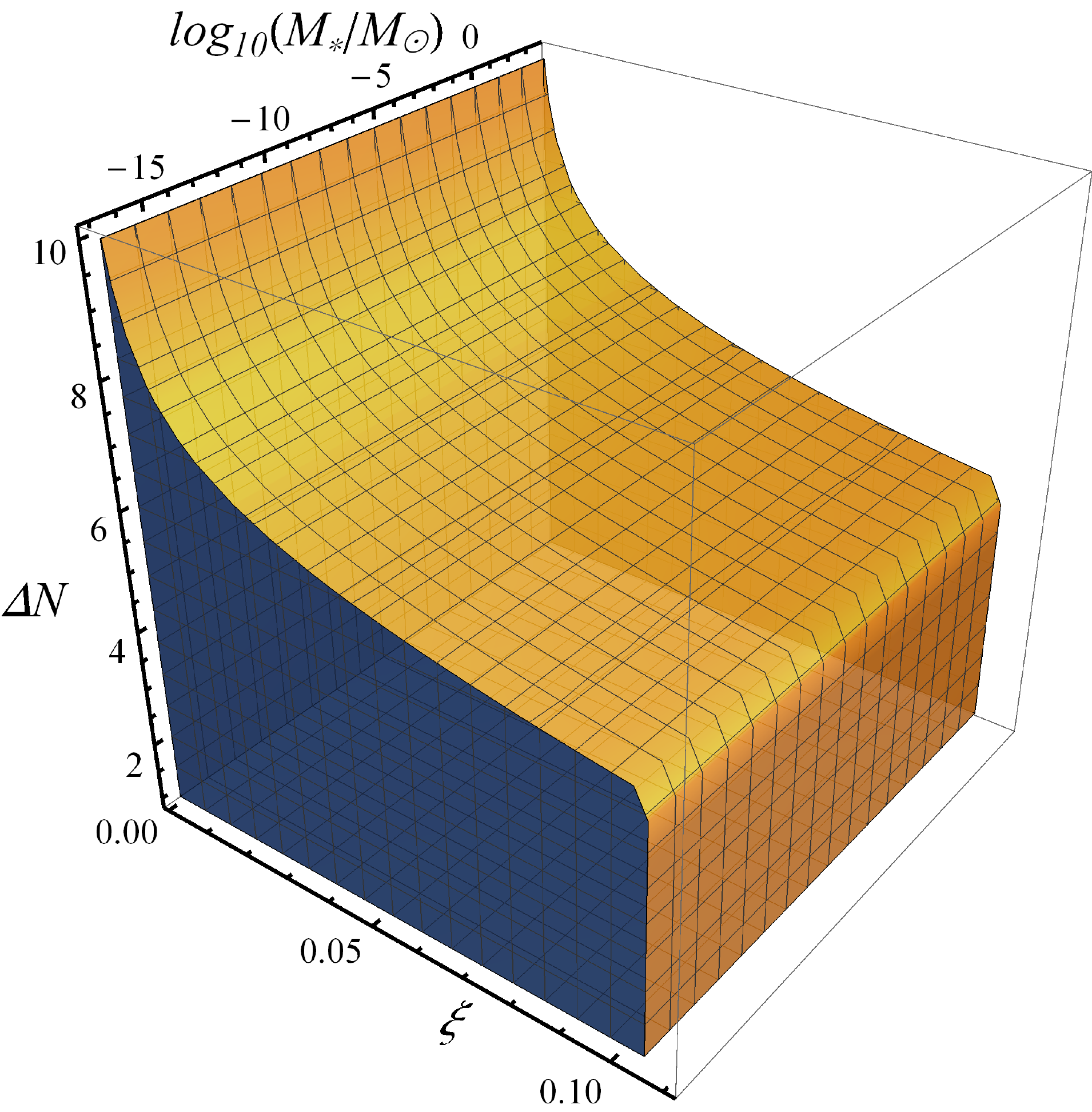}
	\caption{The constraints on the oscillation amplitude $\xi$ from various astronomical experiments as shown in Fig. \ref{fig_fpbh}.}
	\label{fig_xi}
\end{figure}

\begin{figure}[tbp]
	\centering
	\includegraphics[width=3.6in]{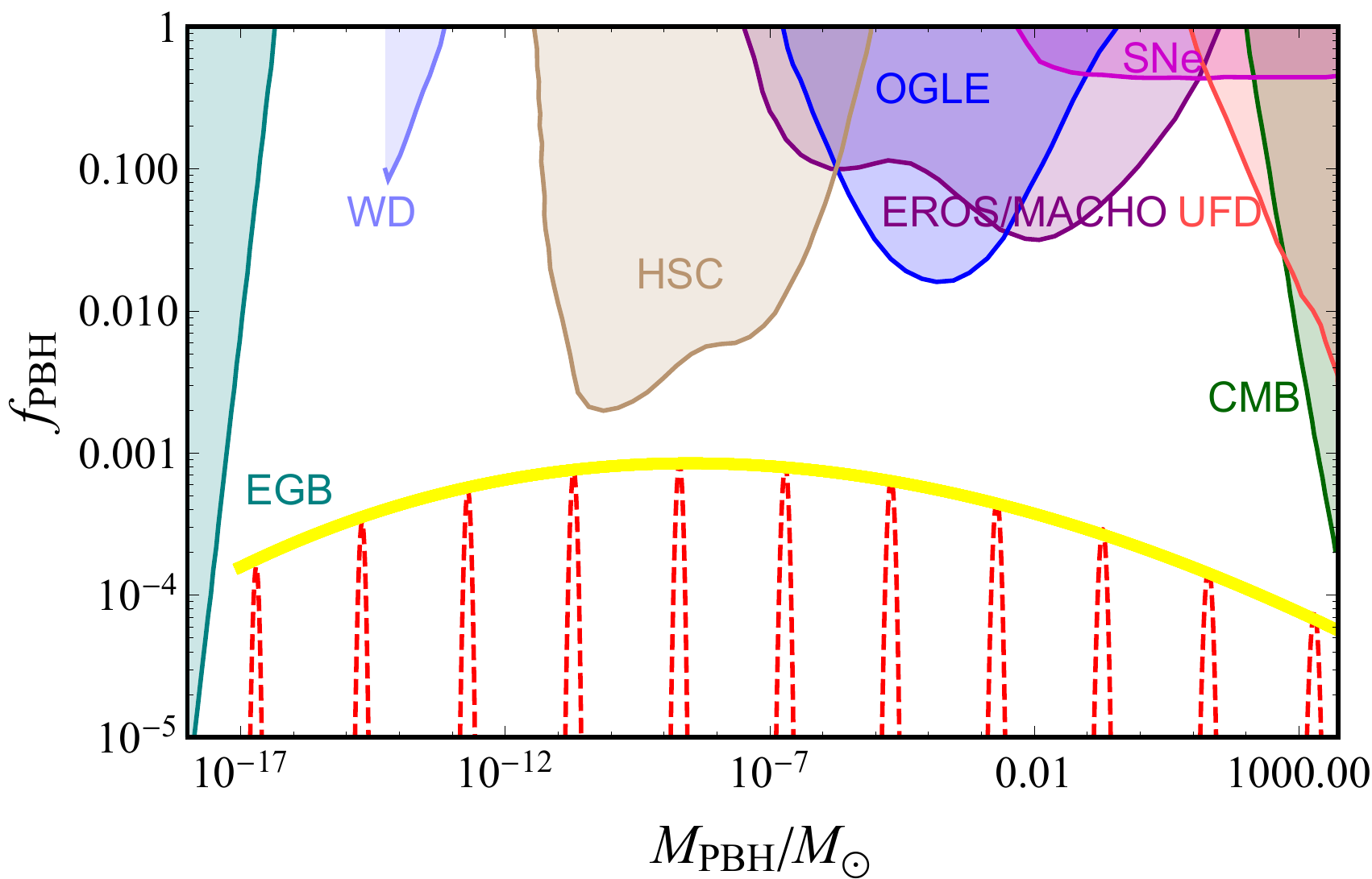}
	\caption{The fraction of PBH against the total DM density, $f_\text{PBH}$, in the inflaton-curvaton mixed scenario, for different values of $k_*$. The colored shadow areas refer to the constraints from various astronomical experiments, including observations of the extra-galactic $\gamma$-ray background (EG$\gamma$), constraints from white dwarves (WD), lensing events (HSC, EROS/MACHO, SNe, OGLE), ultra-faint dwarf galaxies (UFD) and the cosmic microwave background (CMB). We take the value of the parameters as follows: $\xi=0.1$, $\gamma \simeq 0.2$, $g_{*,\text{form}} = 100$, $\delta_c = 0.37$ and $n_s = 0.968$.}
	\label{fig_fpbh}
\end{figure}

To estimate the abundance of PBHs with mass $M$, one usually defines $\beta(M)$ as the mass fraction of PBHs against the total energy density at the formation, which can be expressed as an integration of the Gaussian distribution of the perturbations
\be
\beta(M) \equiv \frac{\rho_\text{PBH}(M)}{\rho_\text{tot}} = \frac{\gamma}{2} \text{Erfc}\l[ \frac{\delta_c}{\sqrt{2} \sigma_M} \r] ~,
\ee
where Erfc denotes the complementary error function. $\sigma_M$ is the standard deviation of the density perturbations at the scale associated to the PBH mass $M$, which can be further expressed as $\sigma_M^2 = \int^\infty_0 (dk/k) W(k/k_M)^2 \frac{16}{81} (k/k_M)^4 \mathcal{P}_{\zeta}(k)$, where $W(x) = \exp(- x^2/2)$ is a Gaussian window function. Since the scale-invariant part of the power spectrum is smaller than the critical density, no black holes would form except at scales very near the resonance peak. As we are working in the perturbative regime, the height of the peak in $\mathcal{P}_{\zeta}(k)$ should not exceed unity, corresponding to a maximal variance $\sigma_M^2 \lesssim \frac{8}{81} \xi (k_c/k_p)^{n_s -1} (k_c/k_M)^4 e^{-k_c^2 / k_M^2}$, which is similar with \cite{Cai:2018tuh}. Combining the current experimental constraints shown in Fig. \ref{fig_fpbh}, this maximum variance gives upper bounds for oscillation amplitude $\xi$ with different $\Delta N(k_M)$, which is shown in Fig. \ref{fig_xi}.

PBHs formed by SSR mechanism can account for dark matter in wide parameter ranges and easily satisfy experimental bounds. To see this, we consider the fraction of PBHs against the total dark matter component at present time \cite{Sasaki:2018dmp}
\be
	f_\text{PBH}(M) \equiv \frac{\Omega_\text{PBH}}{\Omega_\text{DM}}
	= 2.7 \times 10^8 \l( \frac{\gamma}{0.2} \r)^{1/2} \l( \frac{g_{*,\text{form}}}{10.75} \r)^{-1/4}
	\l( \frac{M}{M_{\odot}} \r)^{-1/2} \beta(M) ~,
\ee
where $g_{*,\text{form}}$ is the total relativistic degrees of freedom at the PBH formation time.

The numerical results of $f_\text{PBH}$ with the current experimental constraints are shown in Fig. \ref{fig_fpbh}. We have chosen the parameters as follows: $\gamma \simeq 0.2$, $g_{*,\text{form}} = 100$ \cite{Carr:2009jm} and the threshold value for PBH formation $\delta_c = 0.37$ \cite{Sasaki:2018dmp, Escriva:2019phb}, as well as adopting the Planck result for $n_s = 0.968$ \cite{Akrami:2018odb}, and choosing the oscillation amplitude to be $\xi = 0.1$ \cite{Cai:2018tuh}. Fig. \ref{fig_fpbh} shows the current bounds of experiments including extra-galactic $\gamma$-ray background (EGB) \cite{Carr:2009jm}, white dwarves (WD) \cite{Graham:2015apa}, and lensing events including Hyper Suprime-Cam (HSC) \cite{Niikura:2017zjd}, Exp\'{e}rience pour la Recherche d'Objets Sombres (EROS) / massive astrophysical compact halo object (MACHO) \cite{Tisserand:2006zx}, Optical Gravitational Lensing Experiment (OGLE) and supernovae (SNe) \cite{Zumalacarregui:2017qqd}. Fig. \ref{fig_fpbh} also shows the constraints from ultra-faint dwarf galaxies (UFD) \cite{Brandt:2016aco} and the CMB background \cite{Poulin:2017bwe, Sato-Polito:2019hws}. In Fig. \ref{fig_fpbh}, the red dashed lines correspond to $f_\text{PBH}$ with different choices for the resonance frequencies $k_*$. Since we obtain the similar maximal variance in density perturbation on the mass scale $M_\text{PBH}$ before, the results for $f_\text{PBH}$ in the mixed scenario are similar with \cite{Cai:2018tuh}. As discussed in \cite{Cai:2018tuh}, one can see from Fig. \ref{fig_fpbh} that the PBH mass distribution is given by a narrow peak around $k_*$: this is a distinctive feature of the PBHs formed by SSR mechanism apart from the PBHs formed by other processes, for which the mass distribution is usually more spread out. By varying the value of $k_*$, the peaks form a 1-parameter family enveloped by a yellow solid curve that mainly depends on the amplitude $\xi$.

\section{Conclusion}
\label{sec:concl}

In this work, we study SSR mechanism in the inflaton-curvaton mixed model, in which the total primordial density perturbation is contributed by both the inflaton and the curvaton fluctuations. In contrast to the original SSR mechanism \cite{Cai:2018tuh}, we assume that the sound speed of the curvaton fluctuations is oscillating instead of the inflaton fluctuations during inflation, which in turn leads to the standard adiabatic perturbation in our model. For simplicity, we consider a quadratic potential for curvaton field and a general inflaton potential characterized by the slow-roll parameter $\epsilon$, which is assumed to be time-independent in the quasi-de Sitter approximation for background. We start from the Mukhanov-Sasaki equation for mode function of the curvaton fluctuations, and it can be rewritten as a standard Mathieu equation in the quasi-de Sitter approximation, which has a narrow resonance range in the model parameter space. 
We have also examined carefully the de Sitter approximation and the quasi-de Sitter approximation in SSR mechanism during inflation, and it turns out that these two approximations match reasonably well with the numerical results for both the resonating inflaton and the resonating curvaton.
Consequently, we find that the mode function is exponentially amplified around the characteristic frequency $k_c$, while the other modes behave as the BD vacuum state. After the inflationary era, inflaton is assumed to decay into radiation immediately, while the curvaton energy density is still subdominant and carries the entropy perturbation which could convert into adiabatic curvature perturbation before curvaton decays. Finally, we obtain the total primordial curvature perturbation which is enhanced on small scales and remains nearly scale-invariant on large scales. Since we work in the perturbative regime, the maximum value in amplified total primordial spectrum must not exceed than unity. By combining with the observational constraints on the primordial density perturbation, we plot the parameter space $\{f,m,h\}$ of the mixed scenario. Moreover, we calculate the non-Gaussianity and provide constraints on the energy density $f$. For the small energy fraction of curvaton at its decay, the relative contribution of curvaton for primordial density perturbation is much larger than that of the inflaton on both small and large scales, and thus inflaton perturbation could be subdominant in the mixed scenario, which is different from other's works on PBH formation via curvaton mechanism. Moreover, inflaton perturbation could become the dominant contribution in the non-resonant region, only if the energy density of the curvaton becomes dominant before its decay.

After the standard calculations, we acquire the maximal variance of the primordial density perturbation on the mass scale $M_\text{PBH}$, which gives the upper bounds for oscillation amplitude $\xi$ combining with the current bounds of various experiments on the PBHs. Similar with the original SSR mechanism studied in \cite{Cai:2018tuh}, the PBH mass distribution is given by a narrow peak around the characteristic scale $k_*$, and by varying $k_*$, one could acquire a wide-range distributed mass spectra from $10^{-17} M_\odot \sim 10^4 M_\odot$, which are within the current bounds from a number of astronomical experiments.

Moreover, we mention that it is interesting to break the degeneracy between the resonating inflaton \cite{Cai:2018tuh} and the resonating curvaton, through the induced GW signals due to different nonlinear couplings between scalar  and tensor modes. The GWs induced by the resonating inflaton from the inflationary era is dominated by the sub-Hubble modes \cite{Cai:2019jah}. When we consider the existence of curvaton, the entropy perturbation arise from curvaton is expected to affect the GW signals induced by the curvaton perturbations. This issue will be addressed in the follow-up study.

\appendix


\section{The solution of the mode function} 
\label{A1A2_app}

The general solution for mode function from \eqref{Mathieu} is given by
\be
\label{Mathieu_Solution}
v_k(\tau) =  A_1 C \Big[ \frac{k^2}{k_*^2} \Big( 1 -2 \xi \Big), \Big( 2 - \frac{k^2}{k_*^2} \Big) \xi, x \Big]  + A_2 S \Big[ \frac{k^2}{k_*^2} \Big(1 -2 \xi \Big), \Big( 2 - \frac{k^2}{k_*^2} \Big) \xi, x \Big] ~,
\ee
where the terms $C[A_k,q,x]$ and $S[A_k,q,x]$ are Mathieu cosine and sine functions, respectively. Moreover, the coefficients $A_1$ and $A_2$ are determined by matching the BD vacuum at the moments earlier than the beginning of the oscillation of sound speed. We assume that at the beginning of oscillation of the sound speed $c_s$, the mode functions $v_k$ stay in the BD vacuum. Thus, we have the joint conditions
\begin{align}
v_k(\tau_i) &= \frac{1}{\sqrt{2 k}} e^{- i k \tau_i} ~, \nonumber\\ 
v_k'(\tau_i) &= \Big( \frac{1}{\sqrt{2 k}} e^{- i k \tau} \Big)' \Big|_{\tau=\tau_i} ~.
\end{align}
The above conditions determine the coefficients $A_1$ and $A_2$ in \eqref{Mathieu_Solution} as follows:
\begin{align}
A_1 &= \frac{ e^{-i k \tau_i} \big( - i k S[A_k,q,-x_i] - k_* S'[A_k,q,-x_i] \big) }{\sqrt{2k} k_* M[A_k,q,-x_i] } ~, \nonumber\\
A_2 &= - \frac{ e^{-i k \tau_i} \big( i k C[A_k,q,-x_i] + k_* C'[A_k,q,-x_i] \big) }{\sqrt{2k} k_* M[A_k,q,-x_i] } ~,
\end{align}
where there is
\be
M[A_k,q,-x_i] \equiv C'[A_k,q,-x_i] S[A_k,q,-x_i] - C[A_k,q,-x_i] S'[A_k,q,-x_i] ~,
\ee
and we have defined $x_i \equiv - k_* \tau_i$. The derivatives of Mathieu functions are denoted by $C'[A,q,x] \equiv \partial C[A,q,x] / \partial x$ and $S'[A,q,x] \equiv \partial S[A,q,x] / \partial x$.


\section{The effective mass term $z''/z$} 
\label{z''/z_app}

The effective mass term $z''/z$ can be expanded in terms of small amplitude $\xi$ up to the first order in the quasi-de Sitter approximation for the resonating inflaton, i.e.
\be \label{z''/z_inflaton}
	\frac{z''}{z} = \frac{2-\epsilon}{(\epsilon - 1)^2 \tau^2} + \frac{4 k_* \xi}{(\epsilon - 1) \tau} \sin(2 k_* \tau)
	+ 4 k_*^2 \xi \cos(2 k_* \tau) + \mathcal{O}(\xi^2) ~.
\ee
Since we focus on the sub-Hubble oscillating modes, i.e. $| k_* \tau_i | \gg 1$, then all the terms containing the slow-roll parameter $\epsilon$ can be dropped in the Mathieu equation \eqref{Mathieu}. Thus, the de Sitter approximation and the quasi-de Sitter approximation in SSR mechanism are similar, which are shown in Fig. \ref{fig_comparison}. We notice that in the de Sitter approximation, $\epsilon \rightarrow 0$, the effective mass term $z''/z$ \eqref{z''/z_inflaton} reduces to the one in \cite{Cai:2018tuh}.

In a general inflation model and a curvaton model (involving a general potential or a non-canonical kinetic term, e.g., involving a coupling to other field(s)), the variable $z$ defined in \eqref{MS_Varible} can be expressed as
\be
z = \frac{f(\tau,\epsilon)}{c_s} ~,
\ee
where $f(\tau,\epsilon)$ is an arbitrary function determined by the dynamics of background curvaton and the evolution of background spactime, the sound speed parameter $c_s$ is given by \eqref{cs}. And then, it is straightforward to expand the effective mass term $z''/z$ in terms of the small amplitude $\xi$:
\be \label{z''/z_general_app}
\frac{z''}{z} = \frac{f''(\tau,\epsilon)}{f(\tau,\epsilon)} + 4 \frac{f'(\tau,\epsilon)}{f(\tau,\epsilon)} \xi k_* \sin(2 k_* \tau) + 4 \xi k_*^2 \cos(2 k_* \tau) + \mathcal{O}(\xi^2)~,
\ee
where the prime denotes the derivatives with respect to the conformal time $\tau$. Note that, the term $4 \xi k_*^2 \cos(2 k_* \tau)$, which arises from the parametrization of the sound speed $c_s$ \eqref{cs}, is dominant on sub-Hubble scales, which is of importance to the resonance effect in Mathieu equation \eqref{Mathieu}, e.g., the resonating curvaton discussed before and the resonating inflation studied in \cite{Cai:2018tuh}. With a consistency check, the general express \eqref{z''/z_general_app} agrees with the results \eqref{z''/z} and \eqref{z''/z_inflaton} in the quasi-de Sitter approximation for the resonating curvaton and the resonating inflaton, respectively.

\acknowledgments

We are grateful to Jie-Wen Chen, Cheng-Jie Fu, Cristiano Germani, Jie Jiang, Stone Pi, Misao Sasaki, Liu Tao, Hong-Tsun Wong, Xi Tong, Dong-Gang Wang, Yi Wang and Sheng-Feng Yan for stimulating discussions. This work is supported in part by the NSFC (Nos. 11722327, 11653002, 11961131007, 11421303), by the CAST Young Elite Scientists Sponsorship (2016QNRC001), by the National Youth Thousand Talents Program of China, and by the Fundamental Research Funds for Central Universities. 
All numerics are operated on the computer clusters {\it LINDA \& JUDY} in the particle cosmology group at USTC.

\end{document}